\documentclass[12pt,twocolumn]{openjournal}

\usepackage{xcolor}
\usepackage{textgreek}
\usepackage[utf8]{inputenc}
\usepackage[english]{babel}
\usepackage{hyperref}
\usepackage{ulem}
\hypersetup{
    colorlinks=true,
    linkcolor=blue,
    filecolor=blue,      
    urlcolor=magenta,
    citecolor=blue,
}
\usepackage{color,colortbl}
\usepackage{tensind}
\tensordelimiter{?}
\DeclareGraphicsExtensions{.bmp,.png,.jpg,.pdf}
\usepackage{verbatim}
\usepackage{orcidlink}
\usepackage{soul}
\usepackage{amsmath}	
\graphicspath{ {./Figures/} }

\begin{document}
\title[Spatially resolved growth]{Self-regulated growth of galaxy sizes along the star-forming main sequence}
\shorttitle{Spatially resolved growth}

\author{Shweta Jain\orcidlink{0009-0009-1792-7199}}
\email{sja281@uky.edu}
\affiliation{Department of Physics and Astronomy, University of Kentucky, 505 Rose Street, Lexington, KY 40506, USA}

\author{Sandro Tacchella \orcidlink{0000-0002-8224-4505}}
\affiliation{Kavli Institute for Cosmology, University of Cambridge, Madingley Road, Cambridge, CB3 0HA, UK}
\affiliation{Cavendish Laboratory, University of Cambridge, 19 JJ Thomson Avenue, Cambridge, CB3 0HE, UK}

\author{Moein Mosleh \orcidlink{0000-0002-4111-2266}}
\affiliation{Biruni Observatory, College of Science, Shiraz University, Shiraz 71946-84795, Iran}
\affiliation{Department of Physics, College of Science, Shiraz University, Shiraz 71946-84795, Iran}

\begin{abstract}
    We present a systematic analysis of the spatially resolved star formation histories (SFHs) using Hubble Space Telescope imaging data of $\sim 997$, intermediate redshifts $0.5 \leq z \leq 2.0$ galaxies from the GOODS-S field, with stellar mass range $9.8 \leq \log \mathrm{M}_{\star}/\mathrm{M}_{\odot} \leq 11.5$. We estimate the SFHs in three spatial regions (central region within the half-mass radii $\mathrm{R}_{50s}$, outskirts between $1-3~\mathrm{R}_{50s}$, and the whole galaxy) using pixel-by-pixel spectral-energy distribution (SED) fitting, assuming exponentially declining tau model in individual pixels. The reconstructed SFHs are then used to derive and compare the physical properties such as specific star-formation rates (sSFRs), mass-weighted ages (t$_{\mathrm{50}}$), and the half-mass radii to get insights on the interplay between the structure and star-formation in galaxies. The correlation of sSFR ratio of the center and outskirts with the distance from the main sequence (MS) indicates that galaxies on the upper envelope of the MS tend to grow outside-in, building up their central regions, while those below the MS grow inside-out, with more active star formation in the outskirts. The findings suggest a self-regulating process in galaxy size growth when they evolve along the MS. Our observations are consistent with galaxies growing their inner bulge and outer disc regions, where they appear to oscillate about the average MS in cycles of central gas compaction, which leads to bulge growth, and subsequent central depletion possibly due to feedback from the starburst, resulting in more star formation towards the outskirts from newly accreted gas.

\end{abstract}


\maketitle

\section{Introduction}
\label{sec:intro}

Galaxy formation and evolution involve a multitude of physical processes operating across various spatial and temporal scales. Despite these diverse pathways, galaxy populations show tight scaling relations across several orders of magnitude, such as the correlation between star formation rates (SFR) and stellar masses ($\mathrm{M}_{\star}$), known as the star-forming main sequence \citep[MS;][]{2004MNRAS.351.1151B, 2007ApJ...670..156D, 2007A&A...468...33E, 2007ApJS..173..267S, 2012ApJ...754L..29W, 2014ApJS..214...15S, 2022ApJ...936..165L, 2023MNRAS.519.1526P}.  
Simulations and (semi)analytical models are calibrated to reproduce the observed MS, suggesting that star-forming galaxies, at any given redshift, tend to self-regulate and grow along the evolving MS, typically with a scatter of not more than $\pm $0.3 dex  \citep{2011MNRAS.415...11D, Lilly_2013, 2013MNRAS.435..999D, 2016A&A...593A..22R, 2016MNRAS.457.2790T, 2019MNRAS.485.4817D, 2023MNRAS.518..456D}. Moreover, the MS of star formation has been observed at least since $z \sim 6$ \citep{2012ApJ...754L..29W, 2017ApJ...840...47B, 2020ApJ...899...58L,2022ApJ...936..165L, 2023MNRAS.519.1526P}.

The propagation of galaxies along the MS can be influenced by various internal and external factors. Galaxies can become positively offset from the MS during periods of enhanced star formation, or starbursts, frequently triggered by events such as galaxy mergers or the inflow of gas \citep{2008MNRAS.384..386C, 2013MNRAS.433L..59P,2014MNRAS.442L..33R, 2023ApJ...950...56H}. Conversely, galaxies can become negatively offset or move off the MS entirely as they undergo quenching processes, significantly reducing or stopping star formation. Quenching can be driven by mechanisms such as feedback from active galactic nuclei (AGN) \citep{1998A&A...331L...1S,2006MNRAS.370..645B, Bluck_2014, Henden_2018}, environmental effects such as the removal of gas due to ram pressure in clusters \citep{1972ApJ...176....1G}, or the rapid consumption of the cold gas reservoir from a star formation burst or violent disc instability \citep{2004ApJ...600..580G, Dekel_2013, 2016ASSL..418..355B}. However, galaxies may only stay quenched permanently if they have reached a certain threshold in halo mass, and the circumgalactic medium (CGM) becomes sufficiently heated to halt cold-mode accretion.  If this threshold is not reached, low-mass galaxies ($10^{7}-10^{9} \mathrm{M}_{\odot}$) may potentially become temporarily quiescent \citep[so-called ``mini-quenching'';][]{2023ApJ...949L..23S,2024Natur.629...53L,2024MNRAS.527.2139D} or massive galaxies may rebuild their discs through gas accretion, eventually returning to the MS, a process known as rejuvenation
\citep{2019MNRAS.489.1265M, 2022MNRAS.513L..10M, 2024PASJ...76....1T}. 

Understanding how galaxies evolve along the MS and eventually migrate to the passive cloud, i.e. how galaxies quench/truncate their star formation, has become one of the key questions in the field of galaxy evolution \citep{2015ApJ...801L..12A,2019MNRAS.487.3845C, 2020MNRAS.497..698T}. To comprehend the various mechanisms that explain the evolutionary fluctuations around the MS, it is essential to have a detailed understanding of the formation and development of the sub-components of galaxies \citep{2003ApJ...582..689M}. Each galaxy type is complex and diverse, composed of elaborate structures and sub-structures. Generally, most galaxies can be approximated by a dense central bulge and/or a disc. Therefore, it is crucial to model galaxies as bulge+disc systems to understand how they evolve \citep{ 2012MNRAS.421.2277L, 2019MNRAS.486..390B, 2024A&A...684A..32J}. The primary disc component is often structurally observed as an extended star-forming component with an exponential surface brightness profile \citep{1977ApJ...217..406K, 1970ApJ...160..811F}.
In contrast, the bulge is a compact central component embedded within the disc, following the $r^{\frac{1}{4}}$ brightness law \citep{1948AnAp...11..247D, 1968adga.book.....S}. Bulges are believed to form and grow in-situ, from early-on star formation, with high central densities already in place by $z = 2$ \citep{2015Sci...348..314T}, followed by further growth through secular evolution \citep[e.g.,][]{2013MNRAS.428..718O}. For example, \citealt{2023arXiv230602472B} reported evidence of an already built-up bulge within a core-disc galaxy (JADESGS+53.18343-27.79097) at a reionization-era redshift of 7.430.

The interplay between the processes shaping these sub-structures and the spatial distribution of star formation can be traced by studying the size-stellar mass relation observed in galaxies \citep{2014ApJ...788...28V, 2017ApJ...838...19W,2023A&A...678A..83M}. Various studies have indicated that high redshift disks increase in size over time, albeit at a slower pace than predicted by simple halo growth models \citep{2010ApJ...721..193P,2016MNRAS.463..832W}. This slower growth may reflect evolving halo spin parameters \citep[e.g.,][]{2008MNRAS.391..481S} or differing impacts of feedback processes \citep[e.g.,][]{2009MNRAS.396..141D}. Observational studies of the properties of galaxies have shown that this inside-out growth is prominent in galaxies across the MS \citep{2015Sci...348..314T,2016ApJ...828...27N, 2022MNRAS.513..256D}. Moreover, at fixed mass, star-forming galaxies
are observed to be larger than their quiescent counterparts \citep{2008ApJ...677L...5V,2013ApJ...773..112C,2014ApJ...788...28V,2024ApJ...960...53V}. Quiescent galaxies, on the other hand, have been observed to grow significantly over time, more than doubling in size from $z \sim 2$ until the present day \citep{2005ApJ...626..680D,2012ApJ...746..162N,2017ApJ...839..127P}. This is consistent with the observation that galaxies were smaller at cosmic noon ($z \sim 1-2$) than they are in the local universe \citep{2015ApJ...799..206B}.  In summary, galaxy sizes do not evolve passively. Star-forming galaxies, after building up their cores, transition toward quiescence through a phase of significant compaction while evolving along the MS \citep{2017ApJ...840...47B, 2017ApJ...837....2M, 2019ApJ...885L..22S}. Essentially, they appear to contract as they quench and become quiescent galaxies, only to expand again as they age \citep{2020ApJ...899L..26S}. However, the underlying mechanisms driving this well-established structural evolution from star-forming to quiescent galaxies remain poorly understood.

In this paper, we analyze the spatially resolved properties of galaxies and their sub-components through their star formation histories (SFHs) across cosmic time, to provide valuable constraints on the galaxy growth and evolution patterns. 
With resolved information, it is possible to discern the quenching patterns within galaxies and elucidate how they position themselves in the SFR-$\mathrm{M}_{\star}$ plane.
In this work, we build upon the spatially resolved analysis presented in \citet{2024MNRAS.527.3291J} to study the SFH of the central regions and outskirts of our targets to shed light on the manner in which the galaxies grow and assemble their masses, across a redshift range of $0.5 \leq z \leq 2.0$ and a stellar mass range of $9.8 \leq \log \mathrm{M}_{\star}/\mathrm{M}_{\odot} \leq 11.5$. We construct the SFHs of galaxies and their sub-components using the spatially resolved pixel-by-pixel spectral-energy distribution (SED) fitting, following the same methodology adopted in \citet{2024MNRAS.527.3291J}. We resolved the galaxies into the ``central'' and ``outskirts'' region on the basis of its half-mass radii. 
The organization of the paper is as follows. In Section~\ref{sec:Method}, we present the data sets and the methodology adopted for the reconstruction of SFHs using pixel-by-pixel SED modeling. In Section~\ref{sec:results} and Section~\ref{sec:Discussion and Conclusions}, we summarize our results, discuss the implications of our study, and present our conclusions.

\section{ Methods } \label{sec:Method}
\subsection{ Data } \label{subsec:Data}
Throughout this work, we use the sample of galaxies with stellar mass range $9.8 \leq \log \mathrm{M}_{\star}/\mathrm{M}_{\odot} \leq 11.5$ up to the redshift $0.5 \leq z \leq 2.0$, taken from \citet{Mosleh_2020}. The upper stellar mass bound and the lower stellar mass bound are motivated by completeness due to volume effects and by completeness due to sensitivity limits, respectively. The upper redshift limit ensures that the galaxies’ rest-frame optical SEDs are probed by several filters, and the lower redshift bound is motivated by completeness due to volume effects. \citet{Mosleh_2020} used the publicly available catalog and imaging dataset, from the 3D-HST Treasury Program \citep{2012ApJS..200...13B,2014ApJS..214...24S} and the Cosmic Assembly Near-IR Deep Extragalactic Legacy Survey (CANDELS; \citealt{2011ApJS..197...35G,2011ApJS..197...36K}) to conduct their study. The 3D-HST catalogs provide the photometry of the sources of all 3D-HST observations and all publicly available data over a wide range of wavelengths (0.3–8 $\mu$m). The PSF-matched mosaic images from the GOODS-South field utilized, including seven filters ($B_{435}, V_{606}, i_{775}, z_{850}, J_{125}, JH_{140}, H_{160}$), due to its extensive filter coverage and best depths, provide the highest S/N ratio on spatially resolved scales for individual galaxies \citep{Guo_2013}. The field covers approximately $170$ arcmin squared. In addition to the 3D-HST filters, ancillary data from other telescopes covering the same field were used to estimate the stellar masses and photometric redshifts (if no spectroscopic or grism redshift is available) of the 3D-HST catalog. These estimates were determined using the \texttt{EAZY} \citep{2008ApJ...686.1503B} and \texttt{FAST} \citep{2009ApJ...700..221K} codes, respectively. 

\begin{figure*}
 \includegraphics[width=\textwidth]{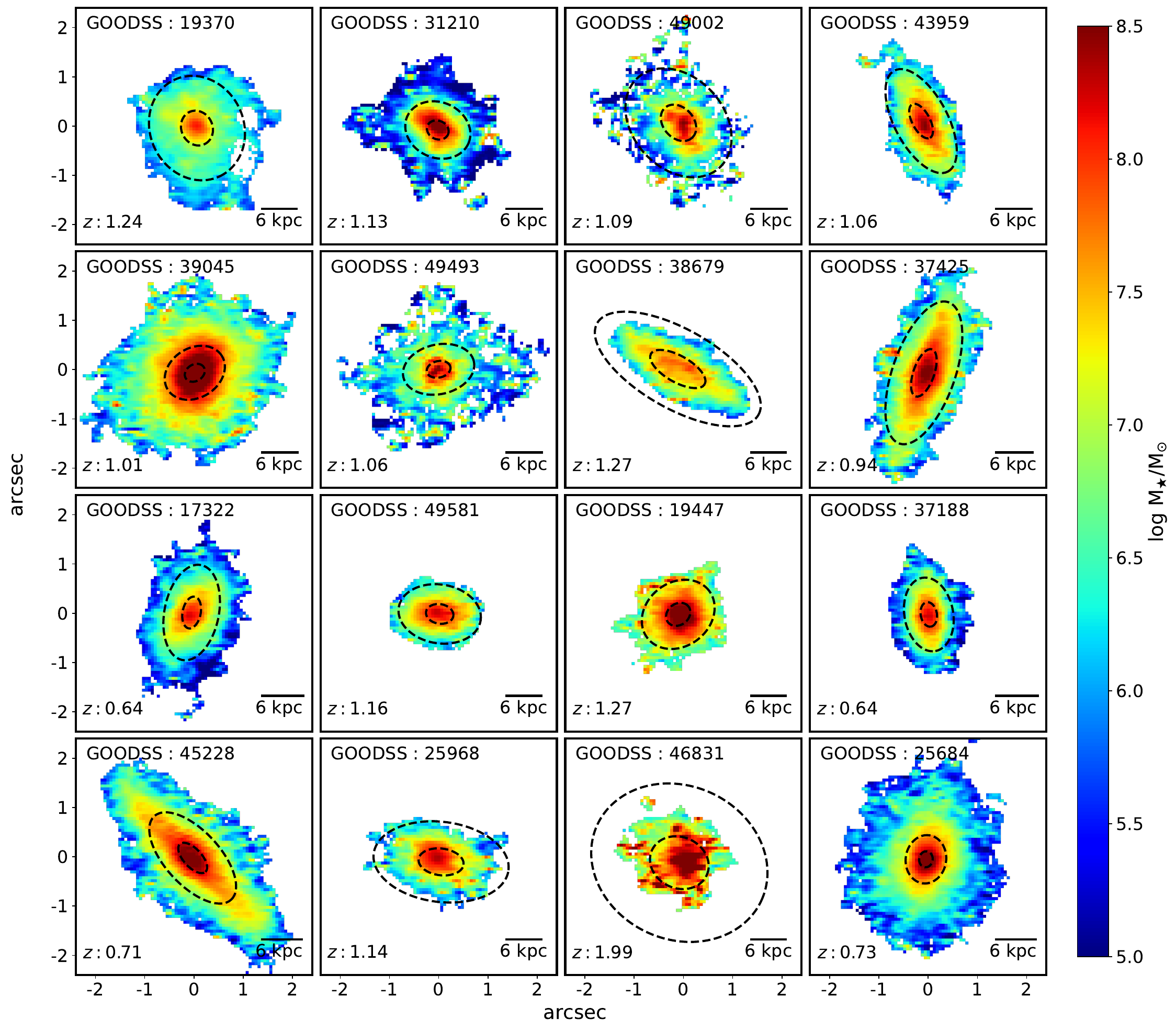}
 \caption{The stellar mass maps (in  $\mathrm{M}_{\odot}$) of galaxies derived using the spatially resolved SED fitting (pixel-by-pixel method; \citealt{Mosleh_2020}). We display the 50$^{\mathrm{th}}$ percentile of the inferred stellar masses. The inner and outer dashed ellipses represent the central and the outskirts regions. The central region encompasses the region within the half-mass radii ($\mathrm{R}_{50s}$) and the outskirts extends from $1-3 \mathrm{R}_{50s}$. The maps are smoothed by a Gaussian kernel of $\sigma = 1$ pixel to improve the clarity of the illustration and enhance the visibility of spatial structure. The photometric redshifts (if no spectroscopic or grism redshift is available) are determined using \texttt{EAZY} code \citep{Brammer_2008} and taken from the HST-catalog \citep{2014ApJS..214...24S}, with corresponding IDs also taken from the same catalog.}
 \label{fig:mass maps}
\end{figure*}

\subsection{ Stellar Population Maps }\label{subsec: 2D}

Spatially resolved physical maps, such as mass maps, age maps, and e-folding timescale ($\tau$) maps on the kpc scale of each of the galaxies, are obtained based on the spatially resolved SED fitting \citep[pixel-by-pixel;][]{Abraham_1999, Conti_2003} method, as described in \citet{Mosleh_2020}. In summary, the authors utilized PSF-matched mosaic images to identify the same physical region of each galaxy in different filters. They selected pixels corresponding to the galaxies using 3D-HST segmentation (mask) maps, with a size of 0.06\,arcsec, corresponding to 0.38 and 0.52\,kpc at $z = 0.5$ and $z = 2.0$, respectively. They extracted fluxes for each pixel in all available filters and determined flux errors using noise-equalized images \citep{2014ApJS..214...24S} from the regions around each object.

The best-fit SED model for each pixel is used to obtain these resolved stellar properties' maps. They performed the SED fitting with \texttt{iSEDfit} \citep{Moustakas_2013}, a Bayesian code, assuming exponentially declining SFHs ($\hbox{SFR} \propto$\,exp(-t$/\tau)$) with the e-folding timescale ($\tau$), between (0.01 -- 1.0 Gyr) and \citet{2003PASP..115..763C} initial mass function (IMF). This exponentially declining model is also known as a simple tau model. The stellar metallicity range used is Z=0.004 -- 0.03, and the \citet{2000ApJ...533..682C} dust attenuation law is assumed. For each galaxy, the redshift of all pixels used is the redshift of the galaxy from the 3D-HST catalog.

We use the same methodology outlined in our previous study \citep{2024MNRAS.527.3291J} to infer the SFHs. In summary, we assume the SFH model to be exponentially-declining ($\hbox{SFR} \propto$\,exp(-t$/\tau$)) and utilize the derived maps of stellar mass, age, and $\tau$ to calculate the SFHs of all pixels within the galaxies. We estimate the SFHs for pixels within 1) the entire galaxy, 2) the central region, and 3) the outskirt region. Finally, we combine the pixels' SFHs by summing the individual pixel-based SFHs to infer the total SFHs of the respective regions. This approach, detailed in equations (1) to (9) of \citet{2024MNRAS.527.3291J}, ensures consistency and includes the propagation of uncertainties.

\begin{figure*}
 \includegraphics[width=\textwidth]{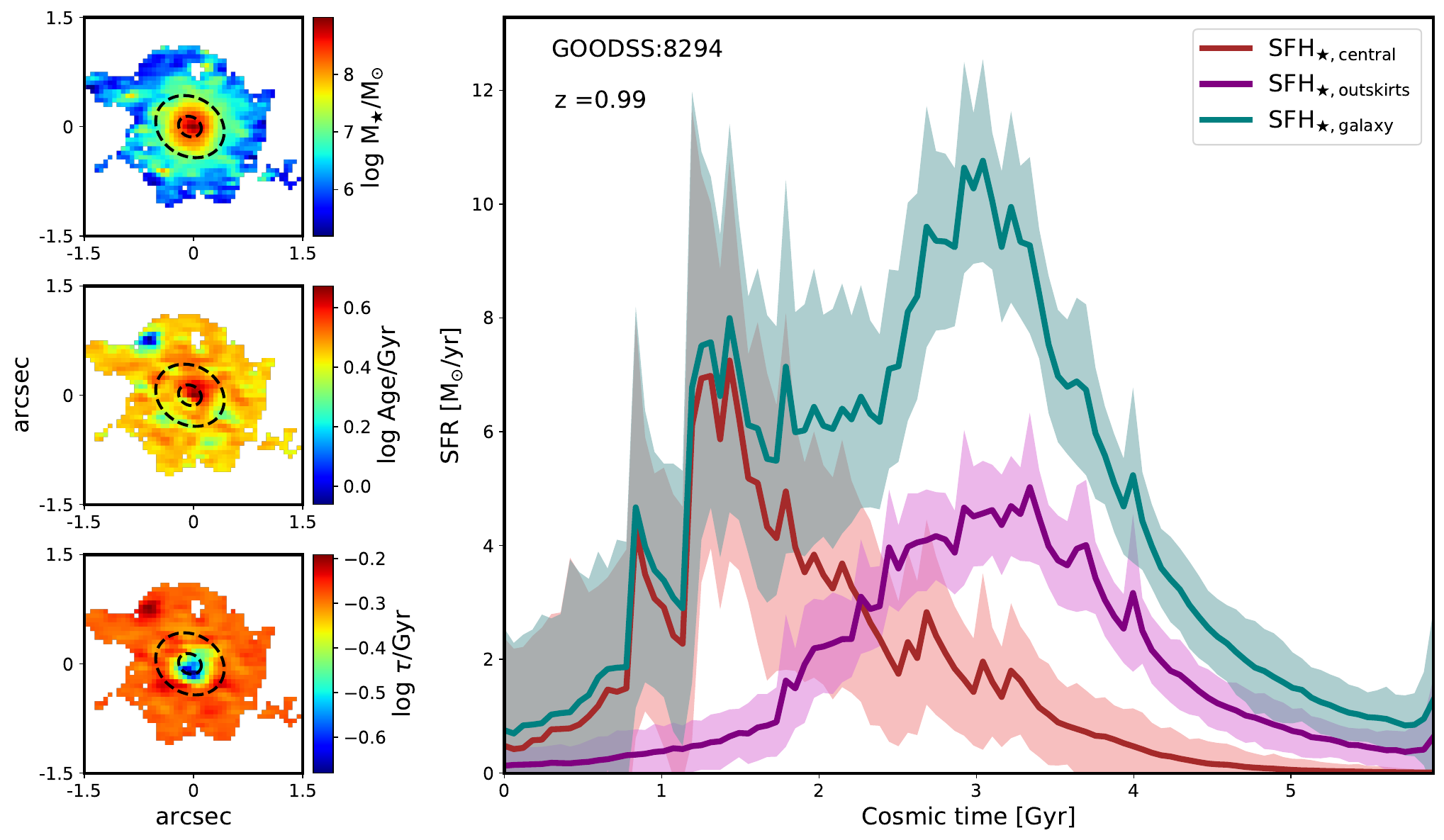}
 \caption{The SFH of an example galaxy. The left panels show the spatially resolved stellar mass map, age map, and $\tau$ map from top to bottom. The inner and outer dashed ellipses represent the central and the outskirts regions, characterized by the region within  $\mathrm{R}_{50s}$ and from $1-3 \mathrm{R}_{50s}$, respectively. The right panel displays the three spatially resolved SFHs obtained from pixel-by-pixel SED fitting for: (\textit{i}) the entire galaxy (green line;  $\mathrm{SFH}_{\star,\mathrm{galaxy}}$), (\textit{ii}) the galaxy's central region (red line;  $\mathrm{SFH}_{\star,\mathrm{central}}$), and (\textit{iii}) the galaxy's outskirts (purple line;  $\mathrm{SFH}_{\star,\mathrm{outskirts}}$). The shaded regions indicate the 16$^{\mathrm{th}}$-84$^{\mathrm{th}}$ percentile ranges. Overall, the depletion in SFR within the central region is followed by the onset of SFR in the outskirts, supporting the inside-out growth pattern within this galaxy.}
 \label{fig:sfhs}
\end{figure*}

\subsection{Central-outskirts division}
\label{subsec: C-O}
We divide the galaxies into two distinct regions: the ``central'' region and the ``outskirts''. For this, we utilize the derived stellar mass maps to determine the half-mass ellipse, the elliptical region which encompasses half of the stellar mass within the galaxy. Using the OpenCV library, we identify the contours within the mass-weighted galaxy maps, with the largest contour representing the galaxy's boundary. This contour is then used to fit an ellipse via OpenCV’s \texttt{fitEllipse} function. During fitting, the semi-major and semi-minor axes ($a$, $b$), along with the orientation angle, are allowed to vary. The fitting process yields key parameters: the center of the ellipse (galaxy's mass-weighted center), the semi-major and semi-minor axes (defining the galaxy’s elongation), and the orientation angle (representing the rotation of the major axis relative to the x-axis). We then fix the orientation angle, vary $a$ and $b$ to define annular elliptical regions centered around the mass-weighted center of the galaxy. We construct the cumulative mass profile using the stellar masses contained within these ellipses and identified the semi-major radius and semi-minor radius corresponding to the half-mass ellipse. We use these radii of the half-mass ellipse ($\mathrm{R}_{50s}$, where the subscript 50 indicates the half-mass and $s$ denotes the elliptical shape of the region) to divide galaxies into the ``central'' and ``outskirts'' region. The ``central'' region is characterized by the region within this $\mathrm{R}_{50s}$. The ``outskirts'' region encompasses the elliptical-annular region beyond the ``central'' region with the outer ellipse extending three times $\mathrm{R}_{50s}$. We use $\mathrm{R}_{50s}$ to divide the galaxy into ``central'' and ``outskirts'' regions to gather information on mass assembly and associated inferred galaxy properties. However, for analyzing the galaxy size–stellar mass relation, we utilize the PSF-corrected circularized half-mass radii, $\mathrm{R}_{50, \mathrm{PSF-corrected}}$ (in kpc), from \citet{Mosleh_2020} to mitigate redshift effects. Specifically, their sizes are circularized using $r_{\mathrm{e,circle}} = r_{\mathrm{e,semi-major}}\sqrt{b/a}$, where $r_{\mathrm{e,semi-major}}$ is the semi-major axis and $\sqrt{b/a}$ is the axis ratio. A potential concern is whether our results depend on the specific size of the aperture used to define the ``central'' and ``outskirts'' regions. To address this concern, we tested various physical regions, such as 1 kpc for the ``central'' and 1-3 kpc for the ``outskirts'', and observed similar trends, shown in our results. The tests confirm that our results are robust and our conclusions remain consistent regardless of whether we use physical or normalized units, or when we vary the definitions of ``central'' and ``outskirts''.

Figure~\ref{fig:mass maps} plots the stellar mass maps (in M$_{\odot}$) of galaxies, derived using the spatially resolved SED fitting. It displays the 50$^{\mathrm{th}}$ percentile of inferred stellar masses. The inner and outer dashed ellipses represent the ``central'' and the ``outskirts'' regions. The maps are smoothed using a Gaussian kernel of $\sigma$ = 1 pixel to enhance the visibility of the spatial structure. 

The SFHs corresponding to each of the defined regions are plotted in Figure~\ref{fig:sfhs}. The left panel of Figure~\ref{fig:sfhs} represents the stellar mass map, age map, and $\tau$ map of an example galaxy from top to bottom. The right panel illustrates the obtained SFH of the ``central'' region (dark red; $\mathrm{SFH}_{\star, \mathrm{central}}$), ``outskirts'' region (purple;  $\mathrm{SFH}_{\star, \mathrm{outskirts}}$) and the whole galaxy (green;  $\mathrm{SFH}_{\star, \mathrm{galaxy}}$). The shaded region indicates the 16$^{\mathrm{th}}$-84$^{\mathrm{th}}$ percentile range in each case. The figure supports the onset of declining SFR within the ``central'' region along with the emergence of SFR within the ``outskirts'', supporting the ``inside-out'' growth pattern in the example galaxy.

\begin{figure}
 \includegraphics[width=\columnwidth]{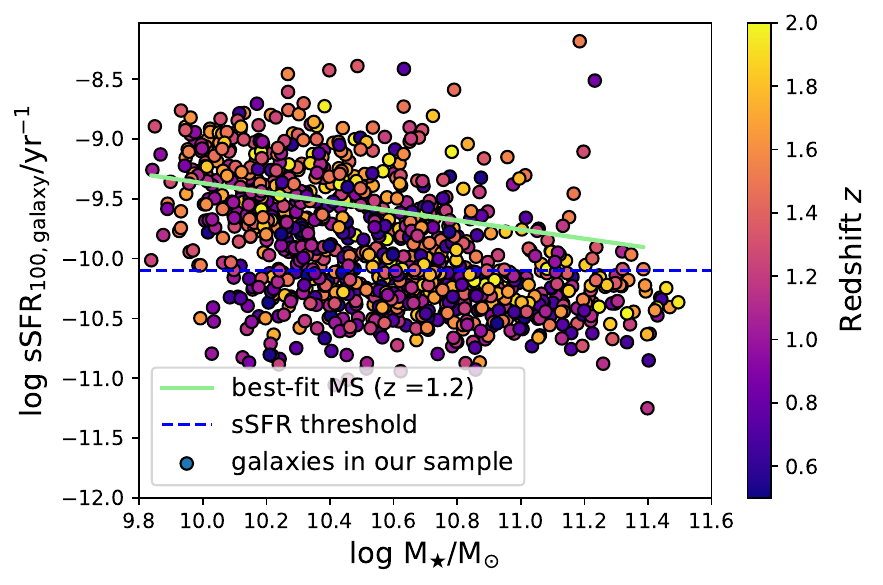}
 \caption{The sSFR-$\mathrm{M}_{\star}$ distribution of the 
 galaxies in our sample (997 galaxies in total), where the sSFR is measured over the last 100 Myr. The distribution is color-coded by the redshift of the galaxies. The spatially resolved SFHs of all the galaxies are employed to infer the stellar masses and sSFRs. The lime green line represents the star-forming MS at redshift of $z = 1.2$ (median redshift of our sample). The MS is defined by the best-fitted MS ridge obtained using Equation~\ref{eq:SFMS}, considering only star-forming galaxies ($\log(\mathrm{sSFR}_{100,\mathrm{galaxy}}/\mathrm{yr}^{-1})>-10.1$; separated by sSFR cut represented by dashed blue line). The stellar mass and redshift limits of our sample are $9.8 \leq \log(\mathrm{M}_{\star}/\mathrm{M}_{\odot}) \leq 11.5$ and $0.5 \leq z \leq 2.0$, respectively.}
 \label{fig:SFMS}
\end{figure}

\subsection{ Defining Star-forming main sequence }\label{subsec: SFMS}
We use the redshift-dependent threshold of specific star formation rate (sSFR) to divide galaxies into star-forming and quiescent \citep{2014ApJS..214...15S, 2016MNRAS.457.2790T, 2022ApJ...936..165L, 2023ApJ...958..183S}.
Observationally establishing a complete absence of star formation presents challenges. As an alternative definition, a galaxy can be considered quenched if its SFR is insufficient to significantly contribute to its stellar mass growth over relevant timescales, such as the Hubble time at the epoch of observation. This relative definition is closely tied to the sSFR, where the inverse of sSFR indicates the time needed to double the stellar mass (accounting for gas recycling). For instance, a galaxy with $\log \mathrm{sSFR}/\mathrm{yr}^{-1} = -10.1$ would require three times the Hubble time at $z=1.5$ to double its mass, indicating it has finished most of its mass assembly and can be considered quenched \citep{2022ApJ...926..134T, Carnall_2019}. To define the MS, galaxies falling below the sSFR threshold will be classified as quiescent, while those above will be considered star-forming. Figure~\ref{fig:SFMS} plots the distribution of galaxies in the  $\mathrm{sSFR}_{100, \mathrm{galaxy}}$-stellar mass plane, where  $\mathrm{sSFR}_{100, \mathrm{galaxy}}$ refers to the sSFR of the whole galaxy measured over last 100 Myr. The plot is color-coded by the redshift of the galaxies. The cut of $\log(\mathrm{sSFR}_{100,\mathrm{galaxy}}/\mathrm{yr}^{-1})=-10.1$ (dashed blue line) divides the galaxies into star-forming ($\log(\mathrm{sSFR}_{100,\mathrm{galaxy}}/\mathrm{yr}^{-1})>-10.1$) and quiescent galaxies ($\log(\mathrm{sSFR}_{100,\mathrm{galaxy}}/\mathrm{yr}^{-1})\leq-10.1$).
It should be noted that this criteria of sSFR threshold to distinguish between star-forming and quiescent galaxies is only used to define the MS. The definition of star-forming MS depends on this criteria, as well as on the chosen parametric model (e.g., power law, broken power law, or more complex parametrization). Consequently, there is an inherent degree of arbitrariness in its definition.
In this study, we define the MS by fitting the distribution of star-forming galaxies ($\log(\mathrm{sSFR}_{100,\mathrm{galaxy}}/\mathrm{yr}^{-1})>-10.1$) in $\mathrm{sSFR}_{100}$-$\mathrm{M}_{\star}$ plane using (Eq.5 of \citealt{ 2016MNRAS.457.2790T}, see also \citealt{2013MNRAS.435..999D}):
\begin{equation}
\label{eq:SFMS}
    \mathrm{sSFR}_{\mathrm{MS}}(\mathrm{M}_{\star},z) = s_{\mathrm{b}}\left(\frac{\mathrm{M}_{\star}}{10^{10}\mathrm{M}_{\odot}}\right)^{\beta} (1+z)^{\mu} \; \mathrm{Gyr}^{-1}
\end{equation}
We find the best-fit values of  $\mathrm{s}_{\mathrm{b}}$, $\beta$ and $\mu$ as $0.22$, $-0.39$ and $0.85$, respectively. We perform Ordinary Least Squares (OLS) linear regression to fit the MS and extract the three variables. The lime-green line  represents the evolution of star-forming MS at redshift of $z = 1.2$ as the median redshift of our sample. As per the calculated MS ridge, our sample galaxies have typically smaller normalization ($\mathrm{s}_{\mathrm{b}}$) at all redshifts when compared to \citet{2014ApJS..214...15S}. Additionally, it has larger stellar mass dependence ($\beta$) and smaller redshift dependence ($\mu$) when compared to \citealt{ 2016MNRAS.457.2790T}.
The deviation can be attributed to different methodologies adopted to calculate both stellar masses and sSFRs of the galaxies, which in our case is largely dependent on the assumed model of SFH. Additionally, we calculated the scatter about the MS ($1 \sigma$ standard deviation) of star-forming galaxies of $\pm 0.35$ dex. In this study, we refer to quiescent and star-forming galaxies based on their distance from the MS or the difference between their sSFR and sSFR$_{\mathrm{MS}}$ ($\Delta_{\mathrm{MS}}$). It should be noted that in the paper, we use the term ``above the MS'' to describe galaxies with $\Delta_{\mathrm{MS}} > 0.0$, and ``below the MS'' to refer to galaxies with $\Delta_{\mathrm{MS}} < 0.0$.

\begin{figure*} 
 \includegraphics[width=\textwidth]{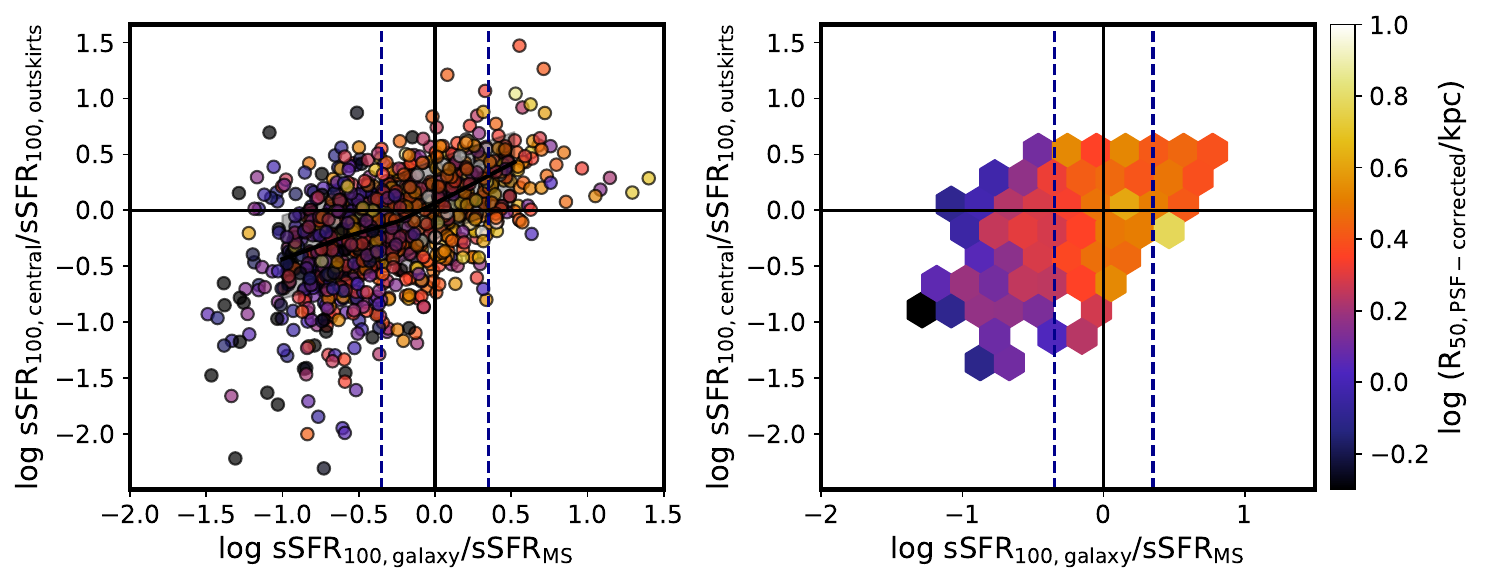}
 \caption{Ratio of the sSFR in the galaxies' central region ( $\mathrm{sSFR}_{100,\mathrm{central}}$; within $\mathrm{R}_{50s}$) to outskirts ( $\mathrm{sSFR}_{100,\mathrm{outskirts}}$; within $1-3 \mathrm{R}_{50s}$) as a function of the distance from the MS ($\Delta_{\mathrm{MS}}$), color-coded by the PSF-corrected half-mass radius ( $\mathrm{R}_{50, \mathrm{PSF-corrected}}$; in kpc). The left panel shows the individual galaxies (997 galaxies), while the right panel shows the average trend by indicating the mean in 2D hex-bin histogram (only regions with 3 or more galaxies are shown). The vertical black solid line represents the $\Delta_{\mathrm{MS}} = 0.0$, while the blue dashed lines show the approximate MS's scatter ($\pm 0.35$ dex). The black solid line with gray shaded region represents the running median in left panel. The key result from this figure is that galaxies at the upper envelope of the MS have positive sSFR ratios -- indicating outside-in growth -- while galaxies at the lower envelope of the MS have negative sSFR ratios -- indicating inside-out growth. Importantly, this is anti-correlated with the sizes of galaxies, implying that larger galaxies typically grow outside-in, while smaller galaxies are growing inside-out. This implies that the growth of galaxies along the MS is self-regulating, explaining why the size-mass relation is rather flat.}
 \label{fig:sSFR_R50_moein}
\end{figure*}

\section{Results}
\label{sec:results}

In this section, we present the results of the various analyses we performed using the SFHs obtained from the spatially resolved scales. Our goal is to understand the growth pattern of galaxies by disentangling the star formation activity within their centers and outskirt regions. We compared the sSFR in the two regions, correlating it with the global galaxy properties such as stellar mass, redshift, distance from the MS (Section~\ref{subsec:sf_distribution}), mass-weighted ages (Section~\ref{subsec:ages}) and half-mass radius (Section~\ref{subsec:size}). While our main analysis focuses on the entire galaxy population (both star-forming and quiescent galaxies), we acknowledge the importance of isolating star-forming galaxies for comparison, particularly as our study emphasizes galaxy evolution along the MS. This is motivated by the potential arbitrariness of the boundary between star-forming and quiescent populations, given that some quiescent galaxies may rejuvenate and rejoin the MS (see Section~\ref{subsec: SFMS} for details). For completeness, we also performed analyses considering only star-forming galaxies, confirming that our results remain robust under this approach. Our results suggest that galaxies tend to exhibit outside-in growth when they are at the upper envelope of the MS (bulge/core building), followed by inside-out growth as they become negatively offset from the MS ridge line (disc formation). These findings are consistent with a picture in which the galaxy pathways along the MS are influenced by major events such as gas compaction, depletion, and quenching. 

\subsection{Star formation distribution within galaxies}
\label{subsec:sf_distribution}

Figure~\ref{fig:sSFR_R50_moein} shows the ratio of sSFR in the galaxies' central regions ($\mathrm{sSFR}_{100, \mathrm{central}}$; within R$_{50s}$), and their outskirts ($\mathrm{sSFR}_{100, \mathrm{outskirts}}$; within $1-3$ R$_{50s}$) as a function of the distance from the MS ($\Delta_{\mathrm{MS}}$; see Section~\ref{sec:Method} for details). The distance from the MS is defined as $\Delta_{\mathrm{MS}}$ $\equiv$ $\log(\mathrm{sSFR}_{100, \mathrm{galaxy}}/ \mathrm{sSFR}_{\mathrm{MS}}$), where $\mathrm{sSFR}_{100, \mathrm{galaxy}}$ represents the sSFR of the whole galaxy measured over 100 Myr. All sSFR values are computed over a timescale of 100 Myr. We plot individual galaxies (total of 997 objects with stellar mass range of $9.8 \leq \log \mathrm{M}_{\star}/\mathrm{M}_{\odot} \leq 11.5$ and redshift range,  $0.5 \leq z \leq 2.0$), in the left panel, while the average trend in a 2D hex-bin histogram is shown on the right. The solid black line marks the running median, with the shaded black region representing 16$^{\mathrm{th}}-84^{\mathrm{th}}$ percentiles. The color coding illustrates the variation with PSF-corrected half-mass radii ($\mathrm{R}_{50, \mathrm{PSF-corrected}}$). Figure~\ref{fig:sSFR_R50_moein} shows a significant correlation between the sSFR ratio on spatially resolved scales with the distance from the MS (see below for partial correlation analysis). Specifically, galaxies above the MS ridge-line ($\Delta_{\rm MS} > 0$) have typically a positive sSFR ratio, implying a higher sSFR in the central region of the galaxy relative to its outskirts. This implies that galaxies double their mass quicker (i.e. on shorter timescales) within the half-mass radius than outside, meaning that the galaxies grow outside-in and their half-mass radius is expected to decrease. 

\begin{figure*}
 \includegraphics[width=\textwidth]{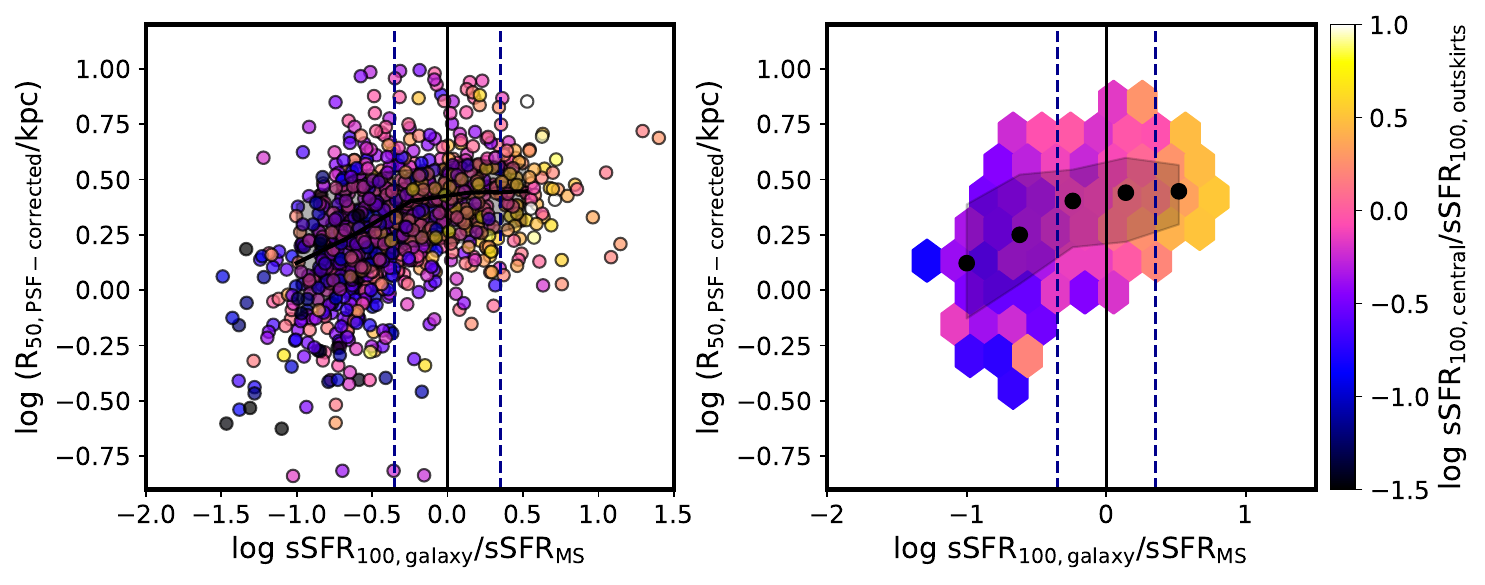}
 \caption{The figure follows the same layout as Figure \ref{fig:sSFR_R50_moein} with y-axis and color-coding exchanged to highlight the half-mass size dependence across the MS. The figure shows clearly the anti-correlation of the galaxy sizes with the sSFR ratio and also the positive correlation with the distance from the MS. Galaxies at the upper envelope of the MS are on average larger, they grow more rapidly their centers than galaxies below the MS. This leads to a self-regulation of the size growth. Galaxies below the MS ($\Delta_{\mathrm{MS}} < -0.35$ dex) that are in the quenching phase are smaller and possibly underwent a compaction phase.}
 \label{fig:sSFR_R50_moein_trend}
\end{figure*}

\begin{figure*}
 \includegraphics[width=\textwidth]{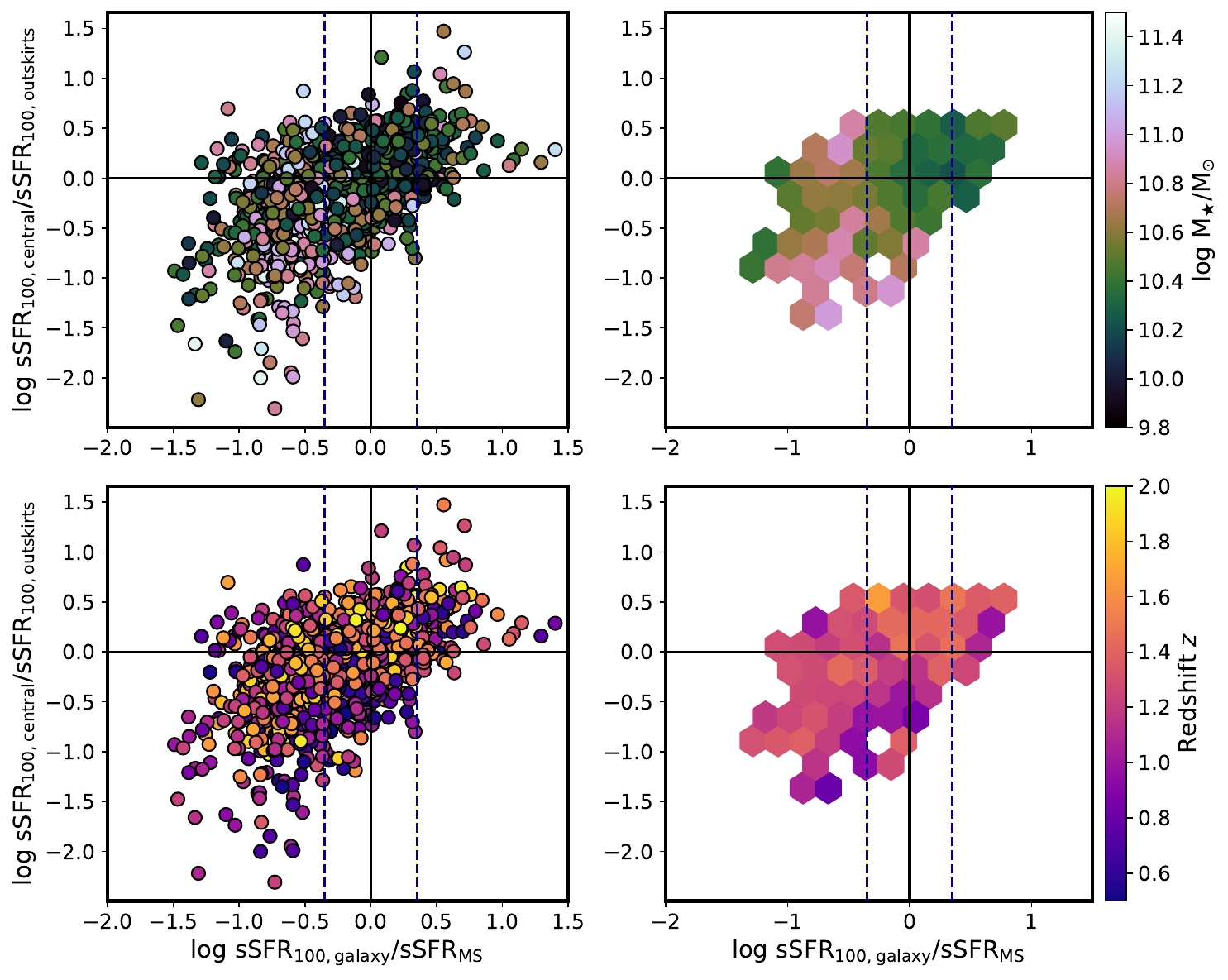}
 \caption{The figure follows the same layout as Figure~\ref{fig:sSFR_R50_moein}. The color-coding illustrates the variation with the total stellar masses (upper panels) and redshifts (lower panels). Most of the higher mass galaxies ($\log \mathrm{M}_{\star}/\mathrm{M}_{\odot} > 10.6$) are towards the lower envelope of the MS ($\Delta_{\mathrm{MS}} < 0$) with low sSFR ratios, implying the depleted sSFR within the central region of galaxies (i.e., inside-out quenching). Lower-mass galaxies ($\log \mathrm{M}_{\star}/\mathrm{M}_{\odot} \leq 10.6$) populate a more uniform distribution around the MS and a range of gradients, implying that these galaxies are growing actively and typically following the trend of outside-in growth above the MS and inside-out growth below the MS. No strong trends can be seen with redshifts, indicating that these conclusions hold over a long period of cosmic time.}
 \label{fig:sSFR_M_z}
\end{figure*}

On the other hand, galaxies toward the lower envelope of the MS ($\Delta_{\mathrm{MS}} < 0$) have a sSFR ratio that is negative, implying faster growth in the outskirts than in their centers, i.e. inside-out growth and an increase in the half-mass radius with time. Many of the galaxies with low sSFR (i.e. $\Delta_{\rm MS} < -0.7$) are massive and are in the process of quenching. This is consistent with previous findings of inside-out quenching \citep{2012ApJ...760..131C,2015Sci...348..314T,2017ApJ...838...19W,2018MNRAS.477.3014B,2022ApJ...935..120J}. 

Importantly, the distance of the MS is also correlated with the half-mass size of the galaxies (color coding in Figure~\ref{fig:sSFR_R50_moein} and Figure \ref{fig:sSFR_R50_moein_trend}): galaxies above the MS are larger in size than galaxies below the MS. This is something that has previously been noted in the literature, where star-forming galaxies are reported to be larger in size than their quiescent counterparts \citep[e.g.,][]{2003MNRAS.343..978S,2008ApJ...677L...5V, 2013ApJ...775..106C, 2014ApJ...788...28V}. This can be seen in Figure~\ref{fig:sSFR_R50_moein}. 
To quantify this correlation, Figure \ref{fig:sSFR_R50_moein_trend} presents the median R$_{50,\mathrm{PSF-corrected}}$ values for galaxies sorted into bins based on the distance from star-forming MS. The results show a clear growth trend: galaxies closer to (within $\Delta_{\rm MS} \pm 0.35$ dex) or above the MS ($\Delta_{\rm MS} > 0.35$ dex) tend to have larger sizes (by $\sim 0.2$ dex than galaxies at $\Delta_{\rm MS} < -0.35$ dex) and show central growth. Conversely, as galaxies move below the MS ($\Delta_{\rm MS} < -0.35$ dex), they become smaller, consistent with a previous compaction event, while their growth currently happens mostly in the outskirts (see Section \ref{sec:intro}).

Therefore, the growth pattern highlighted above (i.e. outside-in growth above the MS and inside-out growth below the MS) is therefore anti-correlated with the half-mass size trend, implying self-regulation for the size growth of galaxies. Specifically, galaxies at the upper envelope of the MS are large, but their outside-in growth leads to smaller sizes, while galaxies at the lower envelope of the MS are small, but their inside-out growth leads to larger sizes. This trend is not unexpected, given that the rather flat size-mass relation is observed. For example, \citet{2019ApJ...877..103S} finds $R_{50} \propto \mathrm{M}_{\star}^{0.1}$ for the star-forming galaxies with $\mathrm{M}_{\star}\approx10^9-10^{11}~\mathrm{M}_{\odot}$, while \citet{Mosleh_2020} finds $R_{50} \propto \mathrm{M}_{\star}^{-0.07}$ for the star-forming galaxies with $\mathrm{M}_{\star} < 10^{10.5}~\mathrm{M}_{\odot}$ and $R_{50} \propto \mathrm{M}_{\star}^{0.2}$ for the star-forming galaxies with $\mathrm{M}_{\star} > 10^{10.5}~\mathrm{M}_{\odot}$. So in summary, the self-regulation along the MS (see Introduction) is also a self-regulation in morphology. 

While the correlation between the spatially resolved sSFR ratio and the distance from the MS is significant (Figure~\ref{fig:sSFR_R50_moein} and Figure~\ref{fig:sSFR_R50_moein_trend}), the question is whether there is another quantity drives this trend. We investigate in Figure~\ref{fig:sSFR_M_z} the dependence of sSFR ratio on stellar mass (upper panel) and redshift (lower panel). The figure follows the same layout as the previous one. The upper panels illustrate that the majority of high-mass galaxies ($\log \mathrm{M}_{\star}/\mathrm{M}_{\odot} > 10.6$) are toward the lower envelope ($\Delta_{\mathrm{MS}} < 0$) with the depleted sSFR within galaxies. These galaxies are growing mostly inside-out with many on their way to quench (i.e., inside-out quenching). While the majority of lower-mass galaxies ($\log \mathrm{M}_{\star}/\mathrm{M}_{\odot} \leq 10.6$) exhibit similar or elevated sSFR in their centers, with an overall increase in sSFR across galaxies.

No such strong trend can be seen with the redshifts of the galaxies, indicating that these conclusions hold over a long period of cosmic time. This is supported by the stronger dependence of spatially resolved sSFR ratio on the distance from the MS than on stellar mass or on redshift. We infer the partial correlation coefficients using the methods by \texttt{PINGOUIN} \citep{Vallat2018} package. We find a partial correlation of $0.48 \pm 0.02$ for the sSFR ratio versus the distance from the MS, $-0.15 \pm 0.03$ for the sSFR ratio versus stellar mass, and $0.18 \pm 0.03$ for the sSFR ratio versus the redshifts of the galaxies.

\begin{figure*}
 \includegraphics[width=\textwidth]{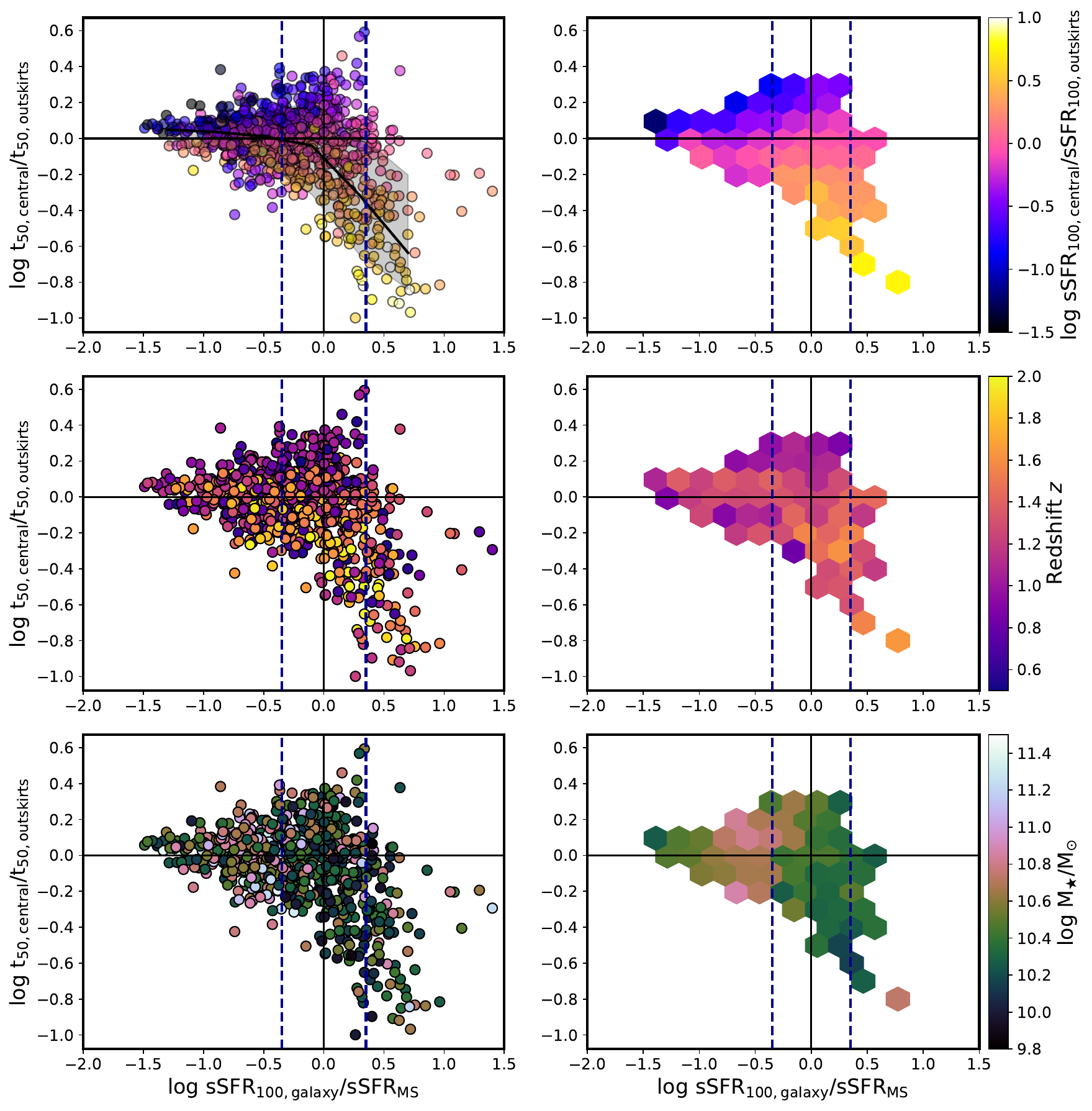}
 \caption{Ratio of the mass-weighted ages of the galaxies' central region ($\mathrm{t}_{50,\mathrm{central}}$) to outskirts ($\mathrm{t}_{50,\mathrm{outskirts}}$) as a function of the distance from the MS ($\Delta_{\mathrm{MS}}$). The ages are in lookback time. The vertical black line represents the $\Delta_{\mathrm{MS}} = 0.0$, while the dashed blue lines show the MS’s scatter ($\pm 0.35$ dex). The color-coding illustrates the variation with the ratio  $\mathrm{sSFR}_{100,\mathrm{central}}/\mathrm{sSFR}_{100,\mathrm{outskirts}}$ (upper panel), redshift (middle panel) and total stellar mass (lower panel). The left panels show the individual data points, while the right panels show the average trend by indicating the mean in hex-bin histograms. The black solid line with gray shaded region represents the running median in the upper left plot. The figure illustrates: (\textit{i}) most of the galaxies above the MS (right envelope; $\Delta_{\mathrm{MS}} > 0$) show younger ages of the central regions ($\mathrm{t}_{50,\mathrm{central}}$), suggesting recent or ongoing mass assembly within their centers, (\textit{ii}) galaxies below the MS (left envelope; $\Delta_{\mathrm{MS}} < 0$) exhibit similar  $\mathrm{t}_{50,\mathrm{central}}$ and  $\mathrm{t}_{50,\mathrm{outskirts}}$, possibly due to the disc-forming central depletion/quenching events within the galaxies, and (\textit{iii}) a few galaxies above the MS (right envelope; $\Delta_{\mathrm{MS}} > 0$) indicate larger  $\mathrm{t}_{50,\mathrm{central}}$ than  $\mathrm{t}_{50,\mathrm{outskirts}}$, suggesting quenched centers and growing discs. These trends align with the color-coded sSFR ratio (upper panel). No such strong trend can be seen with the redshift or stellar mass of the galaxies, indicating that the age ratio is mainly determined by the distance from the MS.}
 \label{fig:t_50}
\end{figure*}

\subsection{Perspective from the mass-weighted ages}
\label{subsec:ages}

The sSFR trends discussed above assess the recent growth, i.e., the growth over a timescale of $\approx100$ Myr. We now expand upon this by looking at mass-weighted ages (t$_{50}$; lookback time when 50\% of the total stellar mass formed), which assess the full SFH and evaluates the growth on longer timescales. Figure~\ref{fig:t_50} shows the ratio of mass-weighted ages of galaxies' center to their outskirts as a function of the distance from the MS. The color coding illustrates the variation with the ratio $\mathrm{sSFR}_{100,\mathrm{central}}/\mathrm{sSFR}_{100,\mathrm{outskirts}}$ (upper panel), redshift (middle panel), and total stellar mass (lower panel). The left panels show the individual data points, while the right panels show the average trend by indicating the mean in hex-bin histogram. 

The plot illustrates an overall trend as follows: \textit{(i)} most of the galaxies with active star formation with respect to the MS (right envelope; $\Delta_{\mathrm{MS}} > 0$) exhibit younger ages of the central regions ($\mathrm{t}_{50,\mathrm{central}}$) than the outskirts regions ($\mathrm{t}_{50,\mathrm{outskirts}}$). This suggests a more recent mass assembly within the galaxies' centers; \textit{(ii)} galaxies below the MS (left envelope; $\Delta_{\mathrm{MS}} < 0$) exhibit similar $\mathrm{t}_{50,\mathrm{central}}$ and $\mathrm{t}_{50,\mathrm{outskirts}}$, possibly due to the onset of quenching of centers along with disc formation; and \textit{(iii)} a few galaxies above the MS (right envelope; $\Delta_{\mathrm{MS}} > 0$) indicate larger $\mathrm{t}_{50,\mathrm{central}}$ than $\mathrm{t}_{50,\mathrm{outskirts}}$, suggesting quenched centers and growing discs. 

The trend aligns with the above findings and the color-coded sSFR ratio (upper panel) where \textit{(i)} centers with enhanced SFR ($\log(\mathrm{sSFR}_{100,\mathrm{central}}/\mathrm{sSFR}_{100,\mathrm{outskirts}})>0.0$) can facilitate ongoing mass assembly within their centers, \textit{(ii)} the SFR depleted centers ($\log(\mathrm{sSFR}_{100,\mathrm{central}}/\mathrm{sSFR}_{100,\mathrm{outskirts}}) < 0.0)$ may suggest the onset of possible ongoing disc-forming central depletion/quenching events within the galaxies. \textit{(iii)} galaxies above the MS ($\Delta_{\mathrm{MS}} > 0$) with the SFR depleted centers ($\log(\mathrm{sSFR}_{100,\mathrm{central}}/\mathrm{sSFR}_{100,\mathrm{outskirts}}) < 0.0)$ may imply quenched centers and growing discs resulting in inside-out growth. No such strong trend can be seen with the redshift or stellar mass of the galaxies, indicating that the age ratio is mainly determined by the distance from the MS.

\subsection{Size evolution of galaxies}\label{subsec:size}
Figure~\ref{fig:R50_m_z} plots the size-mass relation of the galaxies in our sample. The color coding illustrates the variation with the ratio of sSFR (upper panel; $\mathrm{sSFR}_{100,\mathrm{central}}/\mathrm{sSFR}_{100,\mathrm{outskirts}}$) and redshift (lower panel). The left panels show the individual data points, while the right panels show the average trends by indicating the mean in hex-bin histograms. The upper plot shows that the lower mass galaxies ($\log \mathrm{M}_{\star}/\mathrm{M}_{\odot} < 10.6$) that are larger in size tend to have sSFR enhanced centers ($\log (\mathrm{sSFR}_{100,\mathrm{central}}/\mathrm{sSFR}_{100,\mathrm{outskirts}}) > 0.0$), implying that the centers are in the process of assembling mass and decreasing in size (outside-in growth). Conversely, those with smaller half-mass radii exhibit sSFR depleted centers and possibly enhanced sSFR in the outskirts ($\log (\mathrm{sSFR}_{100,\mathrm{central}}/\mathrm{sSFR}_{100,\mathrm{outskirts}}) < 0.0$). Their centers are likely well established and are in the process of forming discs, indicative of an inside-out growth pattern. The smaller half-mass radii may imply contraction of galaxy half-mass size when they start to quench.
Similar to this, galaxies in the stellar mass regime of $\log \mathrm{M}_{\star}/\mathrm{M}_{\odot} \approx 11$ have sSFR depleted centers ($\log (\mathrm{sSFR}_{100,\mathrm{central}}/\mathrm{sSFR}_{100,\mathrm{outskirts}}) < 0.0$) and therefore are growing most effectively inside-out, with many possibly on the way to quench.
On the other hand, no strong conclusions can be drawn from the trend observed with the redshifts of galaxies (lower panel).

\section{Discussion and Conclusions} \label{sec:Discussion and Conclusions}

We present spatially resolved SFHs of $\sim 997$ galaxies in the redshift range, $0.5 \leq z \leq 2.0$, from the GOODS-South field, with stellar mass range $9.8 \leq \log \mathrm{M}_{\star}/\mathrm{M}_{\odot} \leq 11.5$, building up on our previous analysis presented in \citealt{2024MNRAS.527.3291J}. We estimate SFHs in three spatial regions for each galaxy: central region (within $\mathrm{R}_{50s}$), outskirts ($1-3 \mathrm{R}_{50s}$), and the whole galaxy (Fig.~\ref{fig:sfhs}). 

\begin{figure*}
 \includegraphics[width=\textwidth]{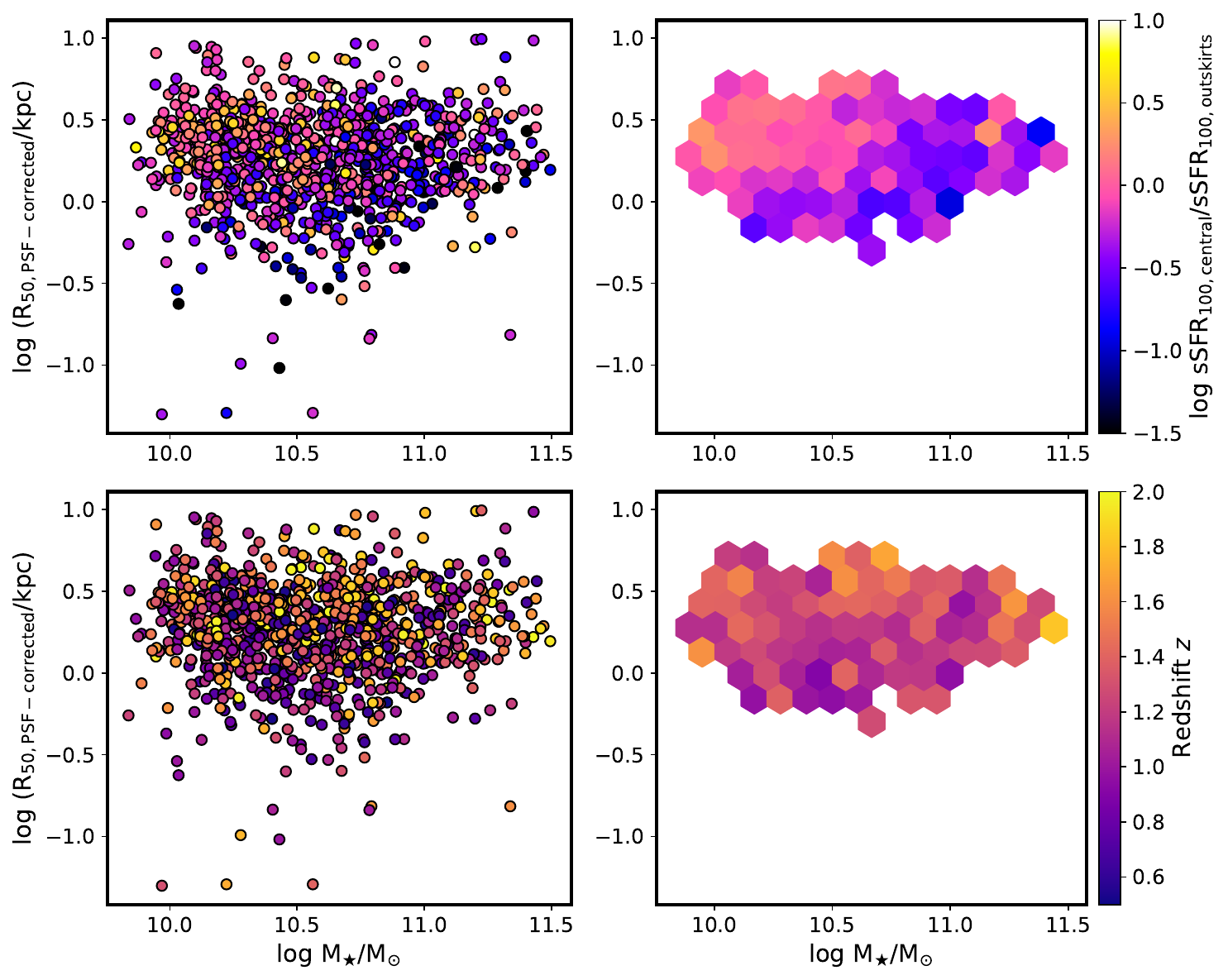}
 \caption{The PSF-corrected half-mass sizes ($\mathrm{R}_{50, \mathrm{PSF-corrected}}$; in kpc) versus stellar masses ($\mathrm{M}_{\star}$; in $\mathrm{M}_{\odot}$), color-coded by the ratio of sSFR (upper panels; $\mathrm{sSFR}_{100,\mathrm{central}}$/$\mathrm{sSFR}_{100,\mathrm{outskirts}}$) and redshifts (lower panels). The left panels show the individual data points, while the right panels show the average trends by indicating the mean in hex-bin histograms.
 The upper plot shows that the small, low-mass galaxies ($\log \mathrm{M}_{\star}/\mathrm{M}_{\odot}< 10.6$) show stronger inside-out growth than larger, low-mass galaxies. Interestingly, the galaxies in the stellar mass regime of $\log \mathrm{M}_{\star}/\mathrm{M}_{\odot} \approx 11$ grow most effectively inside-out, with many in the process of quenching (i.e. inside-out quenching). On the other hand, no such trend can be seen with the redshifts of the galaxies (lower panel).}
 \label{fig:R50_m_z}
\end{figure*}

By correlating the sSFR ratio of the center and outskirts with the distance from the MS, we find that galaxies on the upper envelope of the MS tend to grow outside-in, building up their central regions first, while those below the MS grow inside-out, with more active star formation in the outskirts (Fig.~\ref{fig:sSFR_R50_moein}). While the low-mass galaxies have a preference to grow inside-out, this trend about the MS holds for nearly two orders of magnitude in stellar mass (Figs.~\ref{fig:sSFR_M_z} and \ref{fig:R50_m_z}). Importantly, this growth pattern aligns with galaxy size trends, where larger galaxies above the MS exhibit shrinking sizes due to central growth, and smaller galaxies below the MS grow larger due to active outskirts (Fig.~\ref{fig:sSFR_R50_moein}). Additionally, mass-weighted ages confirm that central regions are younger in star-forming galaxies above the MS, while in galaxies below the MS, the centers and outskirts age similarly (Fig.~\ref{fig:t_50}). The findings suggest a self-regulating process in galaxy size growth when they evolve along the MS. In the following section, we focus on exploring the potential physical mechanisms at play that can lead to this self-regulation.
\subsection{Evolutionary phases about the MS}
\label{subsec: Evolutionary phases}

As outlined in the Introduction, the observed rather small scatter of the MS suggests that galaxies self-regulate their growth and propagate along this SFR-M$_{\star}$ ridge line. Although galaxies' dark matter haloes build hierarchically, most stars form in ``normal'', main-sequence galaxies, which sustain their SFRs for extended periods of time in a quasi-steady state of gas inflow, gas outflow, and gas consumption \citep{2010ApJ...718.1001B,2010ApJ...714L.118D,2012MNRAS.421...98D,2013MNRAS.435..999D,Lilly_2013}. 
Using numerical zoom-in simulations, \citealt[][see also \citealt{2023MNRAS.522.4515L}]{2015MNRAS.450.2327Z} identify four evolutionary phases for galaxies, namely, the ``pre-blue nugget'' phase, i.e., when galaxies are star-forming on the MS; the ``blue nugget'' phase of peak central gas density and SFR; the ``post-blue nugget'' phase characterized by the onset of central quenching; and the quenching phase, where the core has quenched to a ``red nugget''. \citet{2016MNRAS.457.2790T} expanded upon this by showing that the mechanisms of gas compaction, depletion, and replenishment confine star-forming galaxies to the narrow ($\pm$ 0.3 dex) MS. Gas compaction can be driven by a drastic loss of angular momentum due to, e.g. gas rich mergers, counter rotating cold streams, or violent disc instability \citep{2014MNRAS.444.2071D, 2023MNRAS.522.4515L}. As discussed in \citet{2016MNRAS.458..242T}, these phases lead to variations in the sSFR gradient across the MS, with galaxies at the upper envelope of the MS doubling their mass quicker in their cores than their outskirts, and at the lower envelope doubling their mass quicker in their outskirts. This scenario naturally explains why bulges are forming and seen in massive star-forming galaxies on the MS \citep{2014ApJ...788...11L,2015Sci...348..314T}, star-forming galaxies are mostly discs at Cosmic Noon \citep{2009ApJ...706.1364F, 2015ApJ...799..209W}, and galaxies quenching inside-out \citep{2015Sci...348..314T, 2018ApJ...859...56T, 2020ApJ...902...77O, 2021ApJ...915...87S,2023ApJ...943...54J}. Similar pictures have been put forward by other studies of numerical simulations \citep{2021MNRAS.508..219N,2020MNRAS.492.1385K,2022MNRAS.515..213P, 2024MNRAS.527.7871C}.

Our observational findings are consistent with these evolutionary phases across the MS. Specifically, we indeed find that galaxies at the upper envelope of the MS have centrally enhanced star-formation activity ($\log (\mathrm{sSFR}_{100,\mathrm{central}}/\mathrm{sSFR}_{100,\mathrm{outskirts}}) > 0.0$), indicating that they underwent a compaction phase, which fueled the central region with gas. While we do not have information on the detailed kinematics, this central star formation leads to an increase in the central stellar mass density and probably to the formation of bulges. Naturally, this active star formation leads to the depletion of the central gas content through consumption and outflows, marking the onset of a quenching attempt. In the early Universe ($z>3$) and in low-mass galaxies, this could lead to a temporarily quiescent phase \citep[``mini-quenching'';][]{2024MNRAS.527.2139D}. However, in more massive Cosmic Noon galaxies, the depletion times are longer than the replenishment times, leading to fresh gas accretion in the outskirts. We observe the signature of this, where galaxies indeed show a sSFR profile that is rising. This gas, with possibly high angular momentum, could lead to active star formation in a disc like configuration. We indeed confirm that most of the profiles are well reproduced with a S\'ersic index of 1 \citep{Mosleh_2020}. 

This shows that galaxies, while naturally growing along the MS, also self-regulate their growth in size. This is consistent with observed flat size-mass relation at $z=0-3$. The galaxies that depart from the MS and quench do this in an inside-out fashion (rising sSFR profile). It is not clear how long galaxies remain in this state, but since quiescent galaxies have typically smaller sizes than star-forming galaxies, they cannot grow significant amounts of stellar mass at radii greater than half-mass radii. Consistently, it is found that galaxies at higher redshift quench rapidly \citep{2019ApJ...874...17B, 2022ApJ...926..134T, 2024arXiv240417945P}.

\subsection{Limitations and outlook}

Our observations are consistent with galaxies growing their inner bulge and outer disc regions as found in the simulations \citep{2015MNRAS.450.2327Z}, where they appear to oscillate about the average MS in cycles of central gas compaction, which leads to bulge growth, and subsequent central depletion possibly due to feedback from the starburst resulting in more star formation in the outskirts from newly accreted gas. 

In the future, it would be instructive to perform a similar analysis for bulge and disc components separately, in order to study the build-up of bulges more directly. JWST is revolutionizing our understanding of high-redshift galaxies, allowing us to probe lower stellar masses and the rest-frame optical emission to higher redshifts. Extending this analysis to earlier cosmic times, such as $z>3$, would allow us to investigate whether the observed growth patterns and evolutionary pathways of galaxies remain consistent or diverge at epochs closer to the peak of cosmic star formation ($z\sim2$) and beyond, potentially out to $z\sim6$. Such studies could provide deeper insights into the processes driving early galaxy evolution, including bulge formation and quenching mechanisms. 

\section*{Data Availability}
Data available on request.


\bibliography{main}{}
\bibliographystyle{mnras_custom}

\end{document}